# DIGITAL FILTER DESIGNS FOR RECURSIVE FREQUENCY ANALYSIS

HUGH L. KENNEDY

*Defence and Systems Institute, University of South Australia, Mawson Lakes*
*Adelaide, SA 5095, Australia*
*hugh.kennedy@unisa.edu.au*

Digital filters for recursively computing the discrete Fourier transform (DFT) and estimating the frequency spectrum of sampled signals are examined, with an emphasis on magnitude-response and numerical stability. In this tutorial-style treatment, existing recursive techniques are reviewed, explained and compared within a coherent framework; some fresh insights are provided and new enhancements/modifications are proposed. It is shown that the replacement of resonators by (non-recursive) modulators in sliding DFT (SDFT) analyzers with either a finite impulse response (FIR), or an infinite impulse response (IIR), does improve performance somewhat; however stability is not guaranteed, as the cancellation of marginally stable poles by zeros is still involved. The FIR deadbeat observer is shown to be more reliable than the SDFT methods, an IIR variant is presented, and ways of fine-tuning its response are discussed. A novel technique for stabilizing IIR SDFT analyzers with a fading memory, so that *all* poles are inside the unit circle, is also derived. Slepian and sum-of-cosine windows are adapted to improve the frequency responses for the various FIR and IIR DFT methods.

## 1. Introduction

The remarkable efficiency of the fast Fourier transform (FFT) has confined alternative implementations of the discrete Fourier transform (DFT), such as recursive filter banks [1]-[4], to just a handful of niche applications: where input signals are band limited (e.g. in communications [5]-[7] and radar/sonar systems [8]-[10]); where tight architectural constraints are imposed (e.g. in mobile and embedded devices); and/or where low input/output latency is required (e.g. in safety critical [11], closed-loop control and autonomous systems).

In many of the aforementioned digital systems, the DFT is used to form estimates of internal or external system states, via a power density spectrum (PDS) estimate derived from noisy sensor data. On the one hand, when an $M$-point *recursive* DFT is used to form state estimates in a digital feedback-control system, the control loop and the sensor signal digitizer (with an ideal low-pass anti-aliasing filter) are able to operate with a common period of $T$ seconds. On the other hand, when an $M$-point *batch* FFT is used, all $M$ samples must be collected and stored in a buffer before they are processed, thus the controller must operate with a time period of $MT$ seconds between each new control command. This reduces the stability margins of the system, forcing a lower loop-gain to be used, resulting in a 'sluggish' closed-loop system response. The use of even relatively short temporal or spatiotemporal FFTs [12], in guidance/autopilot subsystems of fast-moving autonomous vehicles, where the primary sensor is a video camera operating with a frame rate in the order of 100 Hz, may therefore be sufficient to render the system useless. In many legacy systems however, data are transferred from the sensor to the processor as blocks, rather than as sample-by-sample streams. In these cases, latency is unavoidable and well suited to batch processing using an FFT.

Interest in recursive DFTs for coding, signal analysis and spectrum estimation appears to have increased in recent years [1-11], motivated perhaps by current trends in hardware technology where sampling frequencies of analog-to-digital converters continue to increase, while clock frequencies of digital processors appear to be plateauing after many years of near exponential growth. Sensors and processors are also becoming tightly coupled and being developed as integrated systems, which provides an opportunity for the introduction of new approaches to the management of data flows. There is now, more than ever, a need to develop alternative DFT algorithms that might be more efficient and easier to implement than the FFT on a new generation of sensors and parallel computing platforms [8,10].

In the next section some existing recursive DFT implementations are discussed and compared in terms of their, (impulse- and frequency-) response characteristics and their resistance to the accumulation of rounding errors. Despite the importance of rounding error sensitivity in real systems [1,10,11,13-15]; this point has been overlooked in many studies on recursive frequency analyzers [3-5,16-18] which are highly susceptible to this problem due to the use of pole-zero cancellation on the unit circle. This problem has been addressed for finite impulse response (FIR) Sliding DFTs (SDFTs) in Refs. [13] and [19] through the use of modulators instead of resonators in the filter bank (see Section 2.2). For the most part, this approach does improve performance; however, the dc prototype filter used, still contains a pole on the unit circle at $z = 1$, which leaves the filters vulnerable to rounding errors, when the input signal is corrupted by high-power impulsive noise, for example (see Section 3). It is shown in this paper that this approach may also be applied to the infinite impulse response (IIR) SDFTs (see Section 2.6).

Of the recursive DFT methods considered, the deadbeat-observer technique would appear to be the most attractive [7,20-23], due to its low complexity and rounding error immunity (see Section 2.3); however, it is not ideal for spectrum estimation because its response is fixed and generally unfavorable due to the high side-lobe level of the Dirichlet kernel [24]. Some novel ways of addressing this deficiency are discussed in this paper (see Section 2.4), although freedom to configure and fine-tune this method, by shifting its open-loop poles for instance, is complicated by the outer feedback loop. The fading-memory IIR SDFT methods described in Refs. [16-18], do offer some tuning flexibility; however they *are* susceptible to rounding error accumulation (see Section 2.6). The use of simple band-pass filter banks and exponentially windowed oscillators is also discussed (see Section 2.5).

In addition to providing an introduction and overview of existing recursive methods, with an emphasis on relative strengths, weaknesses and interrelationships, a stabilized version of the fading memory technique described in Ref. [17] is presented here (see Section 2.7). This new filter structure has a recursive frequency analysis stage followed by a non-recursive mixing stage. The optimal mixing coefficients are determined via a least-squares procedure in the time domain but applied as a convolution in the frequency domain to ensure that *all* poles are inside the unit circle. A somewhat less significant, but none-the-less useful, contribution and point of novelty is the adaptation of Slepian windows to FIR frequency-sampling filter-banks (see Section 2.2) and the use of sum-of-cosine windows to improve the response of the various IIR SDFT methods such as Refs. 13. and 17. . To illustrate these main points, results of computer simulations are presented and discussed in Section 3. The paper closes with a summary, some recommendations, and concluding remarks, in Section 4.

## 2. Overview of Some Low-Latency DFT Methods

It is assumed in this paper that the DFT is used to construct a magnitude spectrum estimate $|\hat{X}(n,k)|$, of a one-dimensional signal; therefore filter characteristics that promote the objective of *detection* and *estimation* in the frequency domain are favored (e.g. low variance, low bias, reasonable main-lobe width and low side-lobe levels [22,24]) even if they do not satisfy all the requirements of a formal time-to-frequency *transformation* (e.g. orthogonality and perfect reconstruction [3,21]). As the spectrum is a function of both time and frequency, the methods are well suited to time-frequency analysis [1]; however, the focus here is mainly on steady-state performance for approximately stationary signals. Scope is further limited to non-parametric and non-adaptive methods, implemented using filter banks operating at a common sampling rate. Other than the signal bandwidth and the time duration over which the signal is approximately stationary, no prior information is used to design the filters. The filter responses are time invariant and their maxima are uniformly spaced in frequency. All methods have been factored to conform to a common conceptual architecture (see Fig. 1). This novel decomposition is intended to facilitate the analysis and design comparisons described in this paper.

For notational and coding convenience, *M* is assumed to be odd and all coefficients/signals complex. Using complex variables, in mathematical equations and source code, means that first-order analysis filters may be used; however, nearly twice as many analyzers are required to cover the same frequency range. The associated symmetry benefits are evident when the methods considered here are applied to multi-dimensional signals [12].

Throughout the remainder of this section, the **Methods** implemented in Section 3, involving various permutations of new and existing techniques, are shown in bold typeface. A simple numeric labeling scheme is used to avoid a proliferation of long acronyms. The most basic analysis technique is described in the following subsection. Section 2 is also used: to clarify the nomenclature presented in Fig. 1, to introduce the function of some of the blocks in a familiar context, and to discuss the fundamentals of non-parametric frequency analysis.

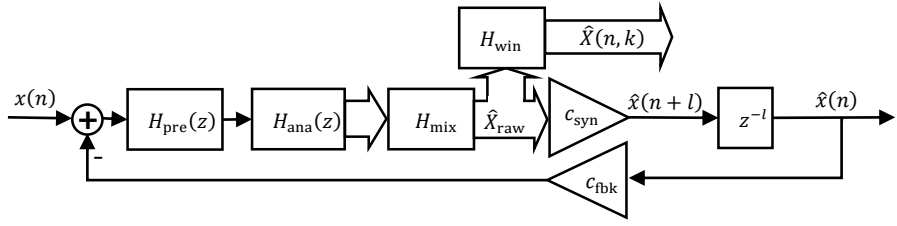

$n$: Discrete-time sample index; $n = 0 \ldots \infty$.

$k$: Frequency 'bin' index.

$K$: Maximum 'measurable' frequency bin index, $-K \leq k \leq +K$.

$B$: Maximum frequency bin of 'interest' for a band-limited signal, $-B \leq k \leq +B$;
    signal bandwidth = $2\,B/M$ (cycles per sample); $B \leq K$.

$M$: 'Nominal length' (in samples) of the odd DFT; $M = 2K + 1$.

$z^{-1}$: Unit delay.

$l$: Prediction horizon (in samples); $l > 0$.

$H(z)$: Discrete-time transfer function.

$c$: Gain factor.

$x(n)$: Digitized input signal.

$\hat{x}(n)$: Estimate of the input signal.

$\hat{X}_{\text{raw}}(n,k)$: Frequency spectrum estimate (raw).

$\hat{X}(n,k)$: Frequency spectrum estimate (windowed).

$H_{\text{pre}}(z)$: Pre-filter; a scalar normalizing gain factor in most cases.

$H_{\text{ana}}(z)$: A parallel bank of frequency-tuned analysis filters.

$H_{\text{mix}}$: Mixing matrix; only used in the stabilized IIR SDFT.

$H_{\text{win}}$: Window function; a convolution in the frequency domain.

$c_{\text{syn}}$: Synthesis factors; the 'partial' IDFT, vector input, scalar output.

$c_{\text{fbk}}$: Feedback gain; unity for FIR/IIR observer, zero otherwise.

Fig. 1. Generic DFT filter architecture.

## 2.1. *FIR DFT*

This non-recursive reference implementation (**Method 1**) is the simplest from a conceptual perspective but the most complex from a computational perspective. The time/frequency elements of the frequency-spectrum estimate $\hat{X}(n,k)$, are scaled bins of the DFT, computed via direct convolution in the time domain using a non-recursive filter with a finite-impulse-response, i.e.

$$\hat{X}(n,k) = \sum_{m=0}^{M-1} w(m) b_k(m) x(n-m) \tag{1}$$

thus

$$H_{\text{ana},k}(z) = \sum_{m=0}^{M-1} w(m) b_k(m) z^{-m} \tag{2}$$

where

$$b_k(m) = e^{j 2\pi m k / M} \tag{3}$$

are the filter coefficients of the frequency analyzer with $j = \sqrt{-1}$. The term 'bin' is used here as a reminder that each filter responds not only to the frequency for which is designed, but also to nearby frequencies.

The pre-filter for this technique simply normalizes the window response using $H_{\text{pre}}(z) = 1/\sum_{m=0}^{M-1} w(m)$. If a rectangular window is applied in the time domain then $w(m) = 1$ for $m = 0 \ldots M - 1$ and $w(m) = 0$ for $m = M \ldots \infty$ thus $H_{\text{pre}}(z) = 1/M$. Evaluating $H_{\text{ana}}(z)$ around the unit circle, i.e. substituting $z = e^{j\omega} = e^{j2\pi f}$ into Eq. (2), yields the frequency response of the $k$th frequency analyzer in the filter bank

$$H_{\text{ana},k}(f) = \frac{\sin\{M\pi(f - k/M)\}}{M\sin\{\pi(f - k/M)\}} \quad (4)$$

which is the Dirichlet kernel $\mathcal{D}_M(f)$ [24], where $f$ is the normalized frequency (cycles per sample) and $\omega$ is the angular frequency (radians per sample).

This response is undesirable because strong but distant signal components all contribute to the output of the $k$th bin, due to the high side-lobes, which may mask the presence of weak but nearby components. Use of a less 'severe/abrupt' or 'tapered' window function, such as the Slepian window (**Method 2**), improves the response by lowering the side-lobe level at the expense of some main-lobe broadening. A broader main-lobe makes it more difficult to resolve strong closely-spaced components, but it also decreases the attenuation of components that are located midway between adjacent bins [24].

The finite impulse response of the Slepian window $w(m)$ for $m = 0 \ldots \infty$, is constructed so that only the first $M$ coefficients are non-zero and so that the proportion of its power density $|W(f)|^2$ within the frequency band $-f_\Delta \leq f \leq +f_\Delta$ is maximized [24]. Convolving the frequency response of the Slepian window with the ideal response of the $k$th frequency analyzer – a unit impulse $\delta(f - k/M)$ – shifts the window response so that the power of the windowed analyzer is maximally concentrated within the band $\frac{k}{M} - f_\Delta \leq f \leq \frac{k}{M} + f_\Delta$. Note that this optimization criterion states nothing about the '*shape*' of the response in the pass band. The response in this band is therefore peaked, not flat; whereas the response outside this band rolls off quickly. Both features are ideal for frequency analysis. Slepian windows, or the first solution of the so-called discrete (or digital) prolate spheroidal sequence (DPSS) were used in this study because the maximal concentration criterion takes some of the guesswork out of the tuning process [25],[26]. For completeness, the procedure for deriving Slepian windows is given in Appendix A. In Ref. 27, the design process has also been extended and generalized for use in digital low-pass filter designs.

As the window is applied in the time domain, $H_{\text{win}}$ and $H_{\text{mix}}$ are simply the identity matrices. The synthesis block applies a 'partial' inverse DFT (IDFT) operation, which reconstructs the noise-free input signal at the $(n + l)$th sample, using only the frequencies within the analysis band $-\frac{B}{M} \leq f \leq +\frac{B}{M}$; when the signal is known to be band limited, $B < K$. This method does not rely on the outer feedback loop to produce a produce the PDS estimate; therefore $c_{\text{fbk}} = 0$ and the synthesis operation is optional; although it may be used if the system is also required to perform a low-pass filtering function. However, the filter will have a very poor response – i.e. poor attenuation outside the pass band and uneven gain with non-linear phase inside the pass band – if a predictive filter is applied ($l > 0$). For best results, assuming a non-zero group delay is tolerable, $l = -K$ should instead be used and the delay block omitted. This yields a linear-phase FIR filter, with a reasonable magnitude response. The synthesis operation is applied using

$$\hat{x}(n + l) = \sum_{k=-B}^{+B} c_{\text{syn}}(k)\hat{X}_{\text{raw}}(n, k) \quad (5)$$

where

$$c_{\text{syn}}(k) = e^{-j2\pi lk/M} \quad (6)$$

and $\hat{x}$ is an estimate of the noise-free signal $\tilde{x}$, see (D.3). Swapping $\hat{X}$ for $\hat{X}_{\text{raw}}$ may be preferable in cases where tapered windows are not required (i.e. for improved frequency resolution).

## 2.2. *FIR SDFT*

The sliding DFT (SDFT) [28],[29], sometimes referred to as the frequency-sampling method in older literature [22], uses a recursive comb pre-filter and a bank of recursive first-order resonator filters with an infinite impulse response (IIR), to generate a bank of frequency analyzers with a finite impulse response (FIR) and a Dirichlet-kernel frequency-response. This FIR SDFT technique is implemented using

$$H_{\text{pre}}(z) = \tfrac{1}{M}(1 - z^{-M}) \quad (7)$$

$$H_{\text{ana},k}(z) = \tfrac{1}{1+a_k z^{-1}} \quad (8)$$

where

$$a_k = -e^{j\omega_k} \text{ with } \omega_k = 2\pi k/M. \quad (9)$$

The outer feedback loop is not utilized by this method.

In Eq. (8) the complex coefficient is in the feedback path of the resonator. This means that the phase reference of the analyzer is always at the current sample ($m = 0$), as per all of the other methods considered in this paper. This configuration is ideal for filtering operations; however, if a fixed phase reference is preferred (e.g. at $n = 0$) then the complex coefficient should instead be placed in the forward path so that it appears in the numerator and denominator of Eq. (8) [29].

Unfortunately, the response of an FIR SDFT analyzer is generated using a marginally-stable pole on the unit circle due to the resonator, cancelled by one of the zeros created by the comb (**Method 3**). This leaves the analyzers vulnerable to the accumulation of rounding errors if the zeros are unable to perfectly cancel the pole due to finite machine precision. This does not mean that the filters are prone to 'explosive' instability; rather, a drift or bias may gradually appear over time. These errors are immaterial if numeric variables are long and if processing intervals are short. The comb pre-filter can be interpreted as being partially responsible for the errors – as it involves the difference of two finite precision numbers. For signals containing only components at the analysis frequencies – i.e. with $M$ being an integer multiple of all component periods – the resulting difference should be zero when the system reaches steady-state, however a non-zero value is output when finite precision is used. The following resonators then integrate these errors, resulting in an offset.

It was shown in Ref. 13. that the stability of the FIR SDFT is improved if resonator convolution – see Eq. (8) – is replaced by a modulator, an integrator (i.e. a pole at $z = 1$), and a phase corrector (the FIR "mSDFT"). The modulating signal is a complex sinusoid which may be generated recursively (**Method 4**) or pre-generated and stored in a look-up table (**Method 5**). Around the same time, a similar approach was proposed and used to transform the temporal dimension of a spatio-temporal filter [19]. The modulating sinusoids *are* pre-generated for this filter; however, complexity is increased through the use of a comb filter in each temporal frequency bin and a non-recursive DFT in the spatial dimensions.

The frequency response of the various SDFT methods may be improved through the application of sum-of-cosine windows in the frequency domain via a convolution operation, which is applied by the $H_{\text{win}}$ block in Fig. 1 [28,30]. These windows (e.g. Hann, Hamming and Blackman) are designed to approximately cancel the side lobes of the Dirichlet kernel using bins that surround the analysis bin (see Appendix C) [24]. However Slepian-type windows may also applied if an 'optimal' approach is preferred; note that *mathematical* optimality is not necessarily the same as *system* optimality, if low side-lobes are more important than narrow main-lobes, for instance. A procedure for deriving the $2B_{\text{win}} + 1$ window coefficients, where $B_{\text{win}} \leq B$, is described in Appendix B. This procedure produces the frequency-domain window coefficients $\widetilde{w}(k)$ for $-B_{\text{win}} \leq k \leq +B_{\text{win}}$, that maximize the proportion of the power density in the frequency band $-f_\Delta \leq f \leq +f_\Delta$, given the constraints of a finite impulse response of $2K + 1$ samples in the time domain *and* a window size of only $2B_{\text{win}} + 1$ bins in the frequency domain. Note that $B_{\text{win}} = 1$ for the Hann and Hamming windows and $B_{\text{win}} = 2$ for the Blackman window. Note also that if $B < K$ then $B_{\text{win}}$ bins at the positive and negative extremes of the spectrum cannot be windowed properly; rather than deriving customized windows for these edge bins, they are left unprocessed here. This Slepian-type window was applied to the FIR mSDFT, implemented using a pre-generated modulator (**Method 6**).

### 2.3. *FIR Deadbeat Observer*

The deadbeat observer is a control-theoretic approach to the problem of frequency analysis (**Method 7**) [7,20-23]. A "deadbeat" response is produced when the output is simply a delayed version of the input, at the sampling times at least. This is achieved by placing all closed-loop poles at the origin of the $z$ plane. An "observer" is a state-space control technique for estimating the internal states of a system, using a plant model embodied within a filter which is driven by a prediction error. The immunity of this method to rounding errors may also be interpreted using a fundamental control theory principle, which asserts that the addition of an integrator to the forward path of a control loop helps to drive steady-state tracking errors of the closed-loop system to zero.

The deadbeat-observer approach to recursive DFT generation employs a bank of integrators modulated by complex sinusoids, to generate a set of equally-spaced marginally-stable poles around the unit circle for the *open-loop* system. A synthesis operation is then performed to yield a one-sample prediction estimate, which is then delayed by one sample, fed back and compared with the input. The bank of frequency analyzers is then driven by the resulting error signal. When the *closed-loop* transfer function of the $k$th frequency analyzer is determined [22.], it can be seen that the poles of the other analyzers in the filter bank form a comb around the unit circle in the $z$ plane (i.e. equally-spaced zeros), one of which cancels the $k$th pole, thus producing the Dirichlet response in the frequency domain.

The observer methods are the only techniques that *need* the feedback loop, in all other cases $c_{\text{fbk}} = 0$. However; all methods may *optionally* use the time-averaged error-signal as an indication of the spectrum quality – i.e. how well the specified frequency components represent the input signal.

### 2.4. *IIR Non-Deadbeat Observer*

The situation described above is perfect for recursive DFT computation – with low complexity and rounding error immunity; however for spectrum estimation, the Dirichlet response is not always ideal. Many years ago when this method was first proposed [20.], Bitmead was quick to point out that the finite-memory (i.e. FIR) deadbeat response of this method may be sub-optimal and that a fading memory (i.e. IIR) might be more desirable in some cases, where greater smoothing (in the time domain) was needed to reduce the effects of random noise [31.]. To deal with this problem he proposed the use of a Kalman filter for frequency analysis [32.]; however, this approach is also not well suited to spectrum estimation as it only reduces the variance of signal components that coincide with the analysis frequencies (narrow main-lobes) and reduces the bias due to the interference of other frequencies (low side-lobes). This approach is more appropriate in applications, such as radar and communication systems, where the component frequencies are known *a priori*.

The *increased* frequency selectivity of a slowly fading memory in a frequency analyzer manifests itself as a narrower/sharper main-lobe with lower side-lobes [16.-18.,32.]. Somewhat counter-intuitively perhaps, window functions in frequency estimation operate by *decreasing* the frequency selectivity of the main lobe. As previously mentioned, this is a side-effect of side-lobe reduction, but is not entirely undesirable as it also increases the response to intra-bin components that might otherwise 'slip through the cracks' and go undetected.

Peceli demonstrated that the response of the deadbeat observer could be improved (low side-lobes without overly narrow main-lobes) by increasing the pole multiplicity of the analyzers in the filter bank to second or third order [22.]. This approach is clearly very attractive; however, in this paper the focus is on banks of first-order IIR filters.

Returning now to Fig. 1: The deadbeat (**Method 7**) and non-deadbeat (**Method 8**) observers are both implemented using: $H_{\text{pre}}(z) = 1/M$; the same analysis filter as the sliding DFT, see Eq. (8); $l = 1$ in Eq. (5) and Eq. (6) and $c_{\text{fbk}} = 1$. Other than increasing the pole multiplicity as discussed above, the observer method is difficult to modify because changes to the various transfer functions have unexpected and usually undesirable consequences on the closed-loop transfer function. The poles must remain on the unit circle, for the reasons discussed above which limits tuning flexibility; however, two simple adjustments are proposed below to modify the response of the deadbeat observer.

The first modification is the use of a band-limited bank of analyzers (i.e. using $B < K$). From a DFT perspective, this involves the use of many samples to evaluate just a few frequency bins. This approach is adopted for all of the other techniques considered in this paper – primarily to reduce the computational load; however, for this particular technique, the closed-loop response of each analyzer is also affected. In the $z$ plane, the poles move radially outwards from the origin thus the system is no longer deadbeat (i.e. FIR). In the *time* domain, the resulting response is less selective as it has a fading memory, with an infinite impulse response (IIR); which increases the *frequency* selectivity of the response. As discussed above, the lowering of the side-lobe levels is particularly useful in frequency analysis applications; however, narrow main-lobes are undesirable when the frequencies of signal components are unknown *a priori*. These effects are most pronounced when $K \gg B$ where the poles approach the unit circle. From a control-system perspective, this change in transient-response behavior in the time domain may be interpreted as being due to the decreased gain applied by the pre-filter in the forward path ($H_{\text{pre}}(z) = 1/M$). Unfortunately this band-limited approach does result in some distortion of the main-lobe maxima (see Section 3).

As has already been mentioned, an overly 'sharp' main-lobe is not necessarily desirable in frequency analysis. To moderate this effect, a second (complementary) modification may be used. If the prediction horizon is increased beyond 1 sample ($l > 1$), then this decreases the frequency selectivity of the observer. As mentioned at the end of Section 2.1, where the IDFT is discussed, when the synthesis sample moves away from the center of the analysis window (where $l = -K$) the response deteriorates; thus we have an alternative mechanism for conferring the main-lobe broadening effects of a window function in an observer framework. If a frequency-domain window function is also applied (see Appendix C), then this must be done *outside* the closed-loop circuit (as shown in Fig. 1) [22]. The existence of the pre-filter gain and the prediction-horizon parameters (see Fig. 1), is hidden in the standard representations of the deadbeat observer used in the (historic and modern) literature on this topic; thus their possible utility as response-tuning mechanisms has been largely overlooked.

**2.5.** *IIR Band-Pass Filters*

Another way to handle the problem of rounding error accumulation described in Section 2.2 is to simply use a bank of first-order damped resonators, with poles inside the unit circle, and to eliminate the comb pre-filter. The pole radius is selected to ensure that there is reasonable overlap between the responses of adjacent filters, so that frequency 'coverage' is reasonably uniform; a non-minimum-phase zero may also be placed on the opposite side of the unit circle for rapid roll-off. Using a near-unity pole radius $r$, yields highly frequency-selective filters. One complex 'half' of a real Goertzel filter is produced when $r = 1$; however, as a consequence of their ever-expanding memory, Goertzel filters require the application of a window to concentrate their impulse response in time [3], using one of the approaches in Ref. 30. for instance, although preferably one which does not involve poles on the unit circle [33]. However, if the poles of the analyzers are inside the unit circle, then time concentration is a natural consequence of the filter's fading memory.

This approach is simply the recursive application of a one-sided exponential window $w(m) = e^{\sigma m}$ (where $\sigma < 0$) to an un-damped resonator. When viewed in this way, it is apparent that a variety of generalized windows with impulse responses of the form $w(m) = m^\kappa e^{\sigma m}$, could be applied to customize the response, if the order of the recursive window is increased using the 'shape' parameter $\kappa$, with $\kappa > 0$ [34.-36.].

### 2.6. *IIR SDFT (Non-Stabilized)*

Like the band-pass filters described in the previous section, the filters described here have an exponentially fading memory; but unlike those analyzers, these analyzers *are* orthogonal, which makes them more suitable for both spectrum sensing and transform functions. They are also particularly useful in communications systems because interference between equally-spaced channels is minimized. This technique features a fading-memory pre-filter in series with a bank of (*un-damped*) resonators [17.]. The comb pre-filter has equally-spaced zeros on the unit circle with frequency-matched poles inside the unit circle. The analyzers are tuned so that their poles cancel one of the zeros on the unit circle, to form the main lobe of the frequency response. Unlike the Dirichlet response of the FIR analyzers (see Sections 2.1, 2.2 & 2.3), the main-lobe of each IIR analyzer is 'sharpened' by the pre-filter's unmatched pole inside the unit circle. In contrast to the deadbeat observer, the pole radius $r$ of the pre-filter is a directly configurable parameter, allowing the time/frequency selectivity to be readily tuned. The IIR SDFT response approaches the FIR response as $r$ approaches zero; furthermore, the response of each analyzer is independent of $B$.

Like the FIR SDFT described in Section 2.2, this IIR SDFT technique also suffers from rounding error accumulation due to pole-zero cancellation on the unit circle – a point which is overlooked in Refs. 16. and 17. . Note that the method described in Ref. 16. is essentially the same as that described in Ref. 17. ; however, each analyzer in Ref. 16. has a zero at $z = 1$, which is apparently added to improve the phase response, and poly-phase extensions are described in Ref. 17. .

This IIR SDFT with a fading-memory (**Method 9**), is implemented using,

$$H_{\text{pre}}(z) = \frac{(1-r)}{M} \cdot \frac{1-z^{-M}}{1-rz^{-M}} \tag{10}$$

and

$$H_{\text{ana},k}(z) = \frac{1}{1+a_k z^{-1}} \tag{11}$$

where

$$a_k = -e^{j\omega_k} \text{ and } r = e^{\sigma M} \tag{12}$$

with $\sigma$ being a forgetting-factor ($\sigma < 0$), which is multiplied here by a factor of $M$ to ensure that the pole radius is the same as the filters described in the next section. This technique does not use the outer feedback loop ($c_{\text{fbk}} = 0$) thus the use of the synthesis factors ($c_{\text{syn}}$) is optional.

As described in Section 2.2 on the recursive DFT with a *finite memory* (the FIR SDFT, **Method 2**), the analysis resonators in this recursive DFT with a *fading memory* (the IIR SDFT, **Method 9**) may also be replaced by (pre-generated) modulators (the IIR mSDFT, **Method 10**). It will be shown in Section 3 that this approach does improve stability somewhat, for both response types, but not in all situations – A fully stabilized version is presented in the next section. Note that in the recent literature the SDFT acronym is usually used to refer to the FIR technique described in Section 2.2; however, in this paper, due to their architectural similarity (different pre-filters), the less well-known IIR technique described in this section (sometimes called the notch Fourier transform [16]) is also referred to here as an SDFT variant.

Like the other DFT versions, application of a window is also optional; although doing this (in the frequency domain) with just a few bins is counterproductive if the poles are close to the unit circle because each main-lobe is very narrow and the side-lobes flat and broad, with sharp notches at the frequencies of the other bins. As a consequence, side-lobes do not cancel effectively as they do for Dirichlet kernels and the windowed main-lobe features multiple maxima rather than a smooth monotonic decay down to the edge of the first side-lobe. If sum-of-cosine windows are to be applied effectively (see Appendix C), then the pole radius $r$, of the pre-filter should be set so that the frequency response is at least somewhat similar to the Dirichlet kernel. A Hann window was applied to the IIR mSDFT, implemented using a pre-generated modulator (**Method 11**).

## 2.7. *Stabilized IIR SDFT*

The problem of rounding-error accumulation that destabilizes the FIR- & IIR-SDFT methods (Sections 2.2 & 2.6) is caused by the filter-bank integration of the rounding errors generated by the comb pre-filter. The technique described here avoids this problem by removing the damped comb pre-filter and replacing it with a frequency-domain 'mixing' operation (**Method 12**). A fading-memory response is instead realized using a bank of *damped* resonators. In effect, the mixing operation orthonormalizes a bank of frequency-shifted 'leaky' integrators which act as band-pass filters. In this regard, the technique is a combination of the techniques described in Sections 2.5 & 2.6 – yielding a stable orthogonal filter bank that is suitable for both estimation *and* transformation functions.

The mixing operation is applied using a $(2B + 1) \times (2B + 1)$ 'mixing matrix' $H_{\text{mix}}$, so that the $k$th windowed frequency bin is a linear combination of all $2B + 1$ analyzer outputs. Clearly, this operation is a significant overhead; however, the extra computational effort is worthwhile in applications where: a one-sample latency is required with bias-free long-term filter operation, if orthonormality is desirable and if band-limited signals are expected ($B \ll K$), to keep the computational complexity to a minimum. The elements of the mixing matrix are derived in the time domain using a weighted least-squares procedure that is described in Appendix D. The procedure fits a basis set of complex sinusoids $\psi_k(m) = e^{j\omega_k m}$, to the input signal $x(n)$, using an exponentially decaying weighting function $w(m) = e^{\sigma m} = r^m$. As shown in the Appendix, the forgetting factor determines the pole radius ($r$) of the analyzers in the filter bank. As the pole radius approaches zero, errors may 'creep' into the mixing matrix due to ill-conditioning, because too-many of the low-frequency basis functions are too similar over the shorter time window. In the absence of numerical errors in the *design* and *filtering* processes, the frequency responses of the stabilized and non-stabilized versions of the IIR SDFT are effectively identical for $|f| < B/K$ (there are no nulls outside this band for the stabilized version). The response of the stabilized IIR SDFT is also somewhat similar to the response of the IIR non-deadbeat observer for all $f$.

This is the only technique that uses the mixing matrix $H_{\text{mix}}$. For this technique $H_{\text{pre}}(z) = 1$, as $H_{\text{mix}}$ replaces the comb and also takes care of normalization. In the filter bank,

$$H_{\text{ana},k}(z) = \frac{1}{1+a_k z^{-1}} \quad (13)$$

where

$$a_k = -e^{\sigma + j\omega_k} = -re^{j\omega_k}. \quad (14)$$

All other blocks are the same as those used in the previous section. Application of a Hann window is optional.

Like all of the IIR methods discussed so far, this method yields very sharp main-lobes as the poles approach the unit circle. This heightened frequency selectivity is ideal for accurate phase and magnitude measurements in noisy environments when signal frequencies are known *a priori* (e.g. active sonar, radar and communication systems). However degraded sensitivity at non-design frequencies, relative to the Dirichlet response of the FIR methods, means that signals at unknown frequencies may go undetected.

### 3. Computer Simulations

#### 3.1. *Computer Code*

Source code to evaluate the various algorithms, using the generic architecture shown in Fig. 1, was built using a C++ compiler and executed on a PC with an i5-3570 CPU and a 64-bit operating system. Single-precision and double-precision versions were instantiated to help reveal the relationship between numeric precision and bias due to rounding error accumulation. The specified numeric type was used only in the filter *application* and error *analysis* stages of execution; whereas, double precision was used to *generate* the simulated input signal and *design* the digital filters for both the single and double cases. No attempt was made to optimize the code at the instruction level, as the emphasis was primarily on filter-response tune-ability and rounding-error susceptibility. A non-negligible execution overhead was incurred through the use of polymorphism, diagnostics, and non in-line functions, which were used to facilitate the development process.

#### 3.2. *Method, Results and Discussion*

Table 1 shows the dependence of the magnitude error $|X(n,k)| - |\hat{X}(n,k)|$, of the various DFT methods as a function of time-scale ($n$) and noise environment, for a given frequency bin. Data streams were processed continuously but divided into segments of $10^6$ samples for simulation and analysis purposes. Two types of zero-mean additive noise were examined: Gaussian-noise with a unity standard deviation, added to *each* sample; and impulsive noise, uniformly distributed over $\pm 1 \times 10^6$, and added to the *first* sample of each segment. The root-mean-squared error (RMSE) over each segment was analyzed in the Gaussian case; the instantaneous error (Err.) was used in all other cases. The input signal was generated using $B + 1$ sinusoidal components with unity magnitude, random phase and frequencies that *coincided* with each of the analysis bins over $0 \leq k \leq B$. Real signals were used in all simulations. Over the timescale investigated, the trends noted below for single-precision were barely noticeable for double precision; therefore only single precision results are shown. Note that windowed variants (**Methods 2**, **6** & **11**) were not considered in this first numerical experiment. The results in Table 1 reveal the following:

- Even in the absence of noise, Err. increases over time by two orders of magnitude for **Methods 3** & **9**, due to rounding error accumulation.
- **Methods 4**, **5** & **10** *do* provide some resistance to error accumulation, in general; however, their estimates *are* corrupted by impulsive noise. **Methods 4** & **5** have similar accuracy.
- **Methods 8** & **12** show no signs of error accumulation and recover well after impulse noise events. **Method 12** is only slightly faster than **Method 1**; **Method 8** is much faster than **Method 1**.
- In the Gaussian noise scenario, the IIR **Methods 9**, **10** & **12** are more accurate than the FIR **Methods 1**, **3**, **4** & **5**, due to their narrow frequency response and long impulse response – see Fig. 2 & Fig. 3. Fig. 2 also shows that there is some asymmetric distortion of the main-lobe of **Method 8**.

In a second set of numerical experiments, the ability of the various DFT methods to detect sinusoidal components of *arbitrary* frequency and magnitude was examined (see Fig. 4). The input signal was the sum of two sinusoidal components with frequencies midway between analysis bins: the first had unity magnitude and $f = 7.5/M$; the second had a magnitude of 0.01 and $f = 17.5/M$. Only double precision was used. The responses in Fig. 2 account for the trends shown in Fig. 4. The results in Fig. 4 suggest the following:

- Tapered windows are required to detect weak components that are close to strong components. In this respect, **Methods 2, 6** & **11** are better than their un-windowed counterparts: **Methods 1, 5** & **10**.
- The side-lobe reduction of **Method 6** is nearly as good as **Method 2**.
- **Method 11** has lower side-lobes and a broader main-lobe than **Methods 2** & **6**.
- Using a smaller $\sigma$ for a larger $r$ in **Methods 10-12** increases frequency selectivity (at the expense of temporal selectivity) and decreases the response to arbitrary signal components, i.e. components that are not near frequency bin centers.
- **Methods 11** & **12** are indistinguishable, except for the last bin, due to numerical errors in the design of the mixing matrix used in **Method 12**.

Note that using $B < K$ transforms the FIR deadbeat observer into an IIR non-deadbeat observer; thus **Method 7** could not be examined in either of the numerical experiments discussed above and only **Method 8** results are presented.

Fig. 3 provides additional insight into the operation of the IIR SDFT methods. The IIR SDFTs have an impulse-response with a 'terrace'-like 'envelope', modulated by a sinusoidal 'carrier'. The stabilized and non-stabilized versions generate the same envelope via different mechanisms (when $B = K$). In the absence of windowing, the magnitude over each step or block of length $M$, is constant and equal to $[(1 - r)/M]\exp(\sigma M n_{\text{blk}})$ for $n_{\text{blk}} = 0 \ldots \infty$, where $n_{\text{blk}}$ is the block index. The quantity $|\hat{X}(n, k)|^2$ produced by these IIR SDFTs may therefore be interpreted as being the $k$th bin in a sliding exponentially-weighted Bartlett periodogram. Non-recursive versions of this frequency estimator have been examined in Refs. 37. and 38. .

Table 1. Comparison of total execution time (s) and magnitude error of DFT techniques for $k = 16$ with $K = 64$ and $B = 32$, in noise-free, Gaussian-noise and impulsive-noise scenarios.

| Method | RMSE Gaussian [a] | Err. Impulsive [b] | Err. No Noise [c] | Err. No Noise [d] | Exec. Time |
|---|---|---|---|---|---|
| **1** | 1.80E-01 | -3.60E-07 | -3.60E-07 | -1.20E-07 | 36 |
| **3** | 1.80E-01 | -3.80E-02 | -1.50E-02 | -3.20E+00 | 5 |
| **4** | 7.80E-01 | -7.80E-03 | 3.60E-07 | 6.00E-07 | 8 |
| **5** | 1.80E-01 | -7.80E-03 | 7.20E-07 | 7.20E-07 | 5 |
| **8[e]** | 1.40E-01 | -1.90E-06 | -1.90E-06 | -9.50E-07 | 5 |
| **9[f]** | 9.20E-02 | -3.60E-02 | -1.50E-02 | -3.20E+00 | 9 |
| **10[f]** | 8.80E-02 | -1.20E-03 | 7.20E-07 | 7.70E-07 | 5 |
| **12[f]** | 8.80E-02 | -1.70E-06 | -9.50E-07 | -1.30E-06 | 28 |

[a] Averaged over $n = 10^6$ to $n = 2 \times 10^6$. [b] At $n = 2 \times 10^6$. [c] At $n = 10^6$. [d] At $n = 10^8$. [e] With $l = 1$. [f] With $\sigma = 1/(2M)$.

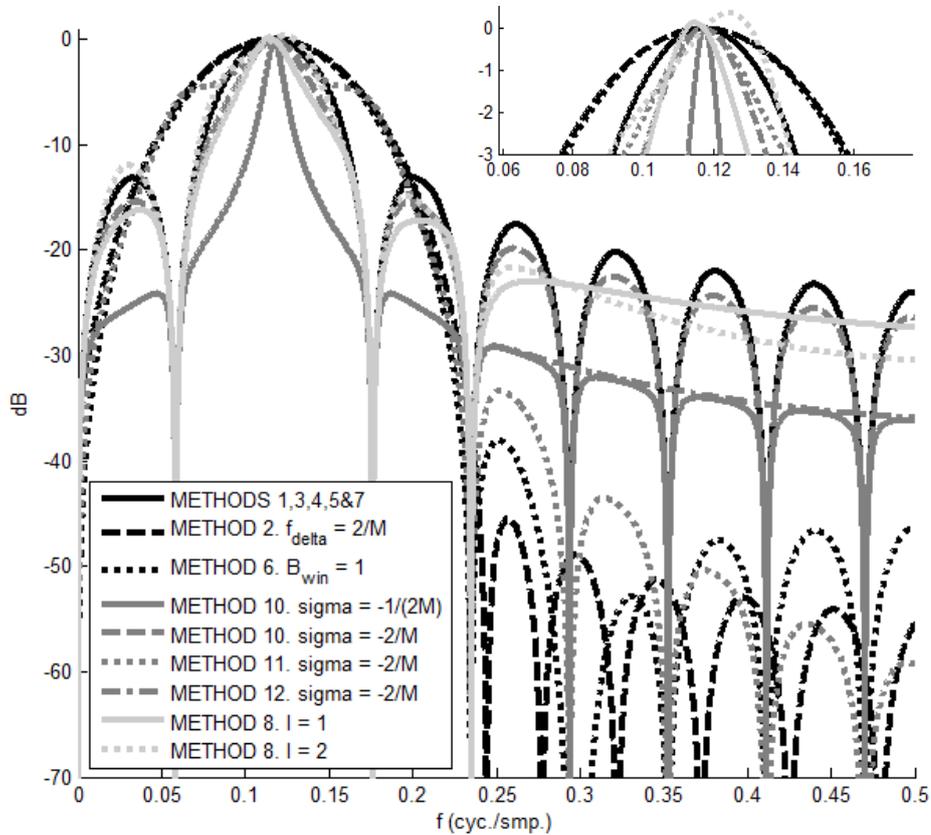

Fig. 2. Ideal frequency responses of DFT filters for $K = 8$, $B = 4$, and $k = 2$. Inset shows detail of main-lobe.

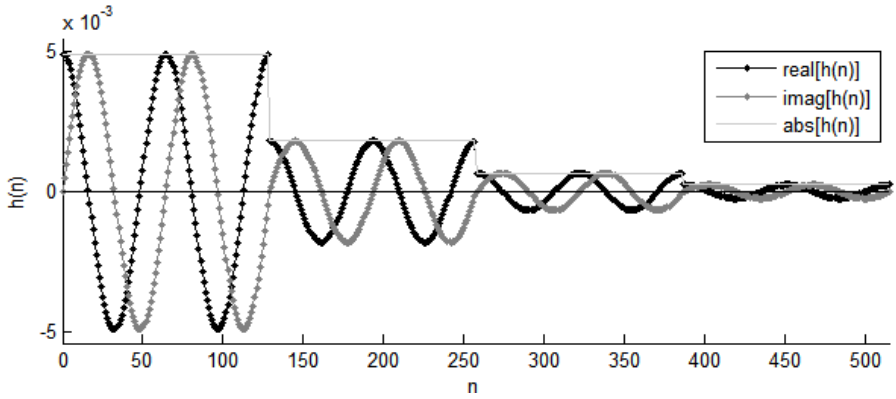

Fig. 3. Impulse response of a non-stabilized IIR SDFT filter (**Method 9**), for $K = 64$ ($M = 129$), $\sigma = -1/M$, and $k = 2$. Application of the Hann window function via a convolution in the frequency domain (**Method 11**) modulates each rectangular block of constant magnitude by the sum-of-cosines taper.

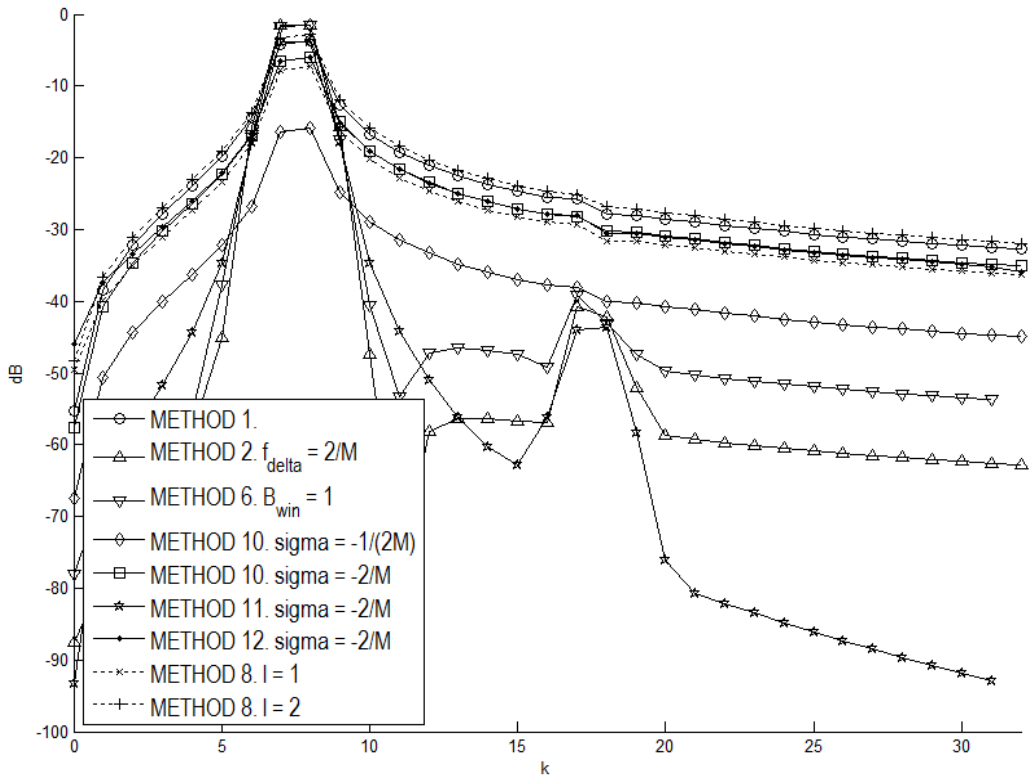

Fig. 4. Detection simulation for $K = 64$ and $B = 32$.

## 4. Conclusion and Summary

### 4.1. *Impulse Response and Filter Structure*

Of the DFT techniques considered here, **Methods 1-7** have an FIR, whereas **Methods 8-12** have an IIR; furthermore, **Methods 1** & **2** are non-recursive, whereas **Methods 3-12** are recursive. **Methods 7** & **8** employ an outer feedback loop. **Methods 2**, **6** & **11** employ tapered window functions. The various filter structures are summarized in Table 2.

Table 2. Filter Structure Summary.

| Method | Class | See Section | Impulse Response | Recursive | Window Function | Outer Feedback loop |
|---|---|---|---|---|---|---|
| 1 | DFT | 2.1 | FIR | No | Rectangular Time | Disabled |
| 2 | DFT | 2.1 | FIR | No | Slepian Time | Disabled |
| 3 | SDFT | 2.2 | FIR | Yes | Rectangular Time | Disabled |
| 4 | mSDFT | 2.2 | FIR | Yes (Modulated) | Rectangular Time | Disabled |
| 5 | mSDFT | 2.2 | FIR | Yes (Pre-Gen. Modulator) | Rectangular Time | Disabled |
| 6 | mSDFT | 2.2 | FIR | Yes | Slepian Frequency | Disabled |
| 7 | Deadbeat Observer | 2.3 | FIR | Yes | Rectangular Time | Enabled |
| 8 | Non-Deadbeat Observer | 2.4 | IIR | Yes | Fading Time | Enabled |
| 9 | mSDFT | 2.6 | IIR | Yes | Fading Time | Disabled |
| 10 | mSDFT | 2.6 | IIR | Yes (Pre-Gen. Modulator) | Fading Time | Disabled |
| 11 | mSDFT | 2.6 | IIR | Yes (Pre-Gen. Modulator) | Fading Time & Hann Freq. | Disabled |
| 12 | SDFT | 2.7 | IIR | Yes (Freq. Mixing) | Fading Time | Disabled |

### 4.2. *Computational Complexity*

The recursive DFT techniques investigated all have the potential to be faster than the FFT for band-limited low-pass signals where filter-banks with $B \ll K$ may be used. Their structure also makes them amenable to parallel implementation. With the exception of the stabilized IIR SDFT (**Method 12**), which is somewhat slower than the other techniques and has a complexity proportional to $(2B + 1)^2$, all of the recursive techniques have complexity that is proportional to $(2B + 1)$, in contrast to the non-recursive FIR DFT (**Method 1**) which has complexity proportional to $(2B + 1)M$. Even when there is no execution-speed advantage to be gained, relative to the FFT which has $M\log_2 M$ complexity, recursive DFT techniques are more appropriate in closed-loop control applications where low measurement latency is desirable.

Unlike all other FIR methods considered, the deadbeat observer technique (**Method 7**) cannot be accelerated by simply reducing the number of frequency bins evaluated, due to the coupling between all analysers via the outer feedback loop; using $B < K$ in the filter bank 'sharpens' the frequency response and 'smears' the impulse response so that it becomes infinite in duration (i.e. IIR), for each closed-loop analyser.

### 4.3. *Numerical Stability*

Computer simulations confirmed that the FIR and IIR SDFTs (**Methods 3-6** and **Methods 9-11**, respectively) are susceptible to rounding error accumulation when single-precision floating-point arithmetic is used. Error accumulation is very gradual for all methods when double precision is used; these techniques may therefore be used with confidence to process short data sets, such as the pixels in a digital image or an audio file. Replacement of resonators by modulators does improve stability somewhat; however, high-power impulsive noise/interference, introduces errors which are never 'forgotten', due to the finite machine precision used in the running sums. This suggests that the FIR and IIR SDFTs proposed so far in the literature are unsuitable in systems intended to handle inputs with a high dynamic range. Even if these anomalous inputs (perhaps due to 'upstream' system exceptions) are extremely unlikely events, their eventual occurrence means that the filter bank, or the entire system, must be reset if it is to continue functioning as designed.

The use of simple band-pass filter-banks with all poles inside the unit circle may be an appropriate solution when only crude indications of channel occupancy are required. However, the proposed stabilized IIR SDFT (**Method 12**), where the filter-bank is orthonormalized in the frequency domain, is a better solution when a more accurate/precise estimator or transformation is required. The outer feedback loop ensures the long-term stability of the (FIR and IIR) observer in all cases examined (**Methods 7** & **8**, respectively).

### 4.4. *Magnitude Error*

For a given frequency bin, using a recursive filter structure (**Methods 3-12**) allows a longer average impulse response duration to be used (i.e. decreased temporal selectivity), without incurring an extra computational cost, which reduces the expected random error in the magnitude estimate due to random noise. Furthermore, the availability of the pole radius parameter ($r = e^\sigma$) in the IIR techniques (**Methods 9-12**) allows the time/frequency selectivity to be fine-tuned for a given combination of $B$ & $K$ paramters, i.e. for a particular distribution of frequency bins. Bias errors due to other frequency components are also reduced, as the effective length of the impulse response increases, due to the narrower main lobe and the lower side lobes. The application of a tapered window function (e.g. Slepian or Hann), in either the time (**Method 2**) or frequency domain (**Methods 6 & 11**), allows biases caused by the side lobes of other 'distant' frequency components to be further reduced, at the expense of a decreased ability to resolve closely-spaced frequency components, due to a broadening of the main-lobe. Of course, the aforementioned benefits are completely lost, however, if rounding errors are permitted to accumulate in recursive implementations. Output characteristics of the various filters are summarized in Table 3.

Table 3. Filter Output Summary.

| **Method** | Main Lobe | Side Lobes | Susceptible to Rounding Error Accumulation |
|---|---|---|---|
| 1 | Narrow | High | No |
| 2 | Wide | Low | No |
| 3 | Narrow | High | Yes |
| 4 | Narrow | High | Yes (In Some Cases) |
| 5 | Narrow | High | Yes (In Some Cases) |
| 6 | Wide | Low | Yes (In Some Cases) |
| 7 | Narrow | High | No |
| 8 | Sharp | Flat | No |
| 9 | Sharp | Flat | Yes |
| 10 | Sharp | Flat | Yes (In Some Cases) |
| 11 | Wide & Non-Monotonic | Low | Yes (In Some Cases) |
| 12 | Sharp | Flat | No |

### 4.5. *Frequency Response Flexibility*

The frequency-domain Slepians were used to reduce the side-lobes of the Dirichlet kernel (**Method 6**). This approach may be applied to the output of any of the recursive techniques where the response is FIR (e.g. the FIR SDFT and the FIR deadbeat observer) or approximately FIR (e.g. the IIR SDFT for small $r$). Sum-of-cosine windows (see **Method 11**) may be applied in the usual way, and for the usual reasons, to the spectral estimates output by all techniques; however, the benefits are limited in IIR cases when the poles are close to the unit circle. The various IIR DFTs (**Methods 8-11)** may be tuned to be very frequency selective (narrow main-lobes with low side-lobes) with good noise rejection and are appropriate when the frequencies of the signal components are known *a priori*, in communications and radar applications, for example. In these cases, the bins may be placed to coincide with the expected component frequencies [16]. Unlike the IIR SDFTs, it is not possible to change the frequency response of the deadbeat observer (**Method 7**) by adjusting the pole positions of the *open-loop* analyzers *directly*; however, the *closed-loop* poles may be shifted *indirectly*, the impulse response changed from FIR to IIR, and the side-lobes of the frequency response lowered, using $B < K$; furthermore, the main-lobes are broadened, for better detection of components with unknown frequencies, by increasing the prediction horizon using $l > 1$.

**Appendix A. Derivation of the Time-Domain Slepian Window**

The discrete-time transfer-function of a non-casual FIR window-function is represented in the $\mathcal{Z}$ domain using

$$W(z) = z^K \sum_{m=0}^{M-1} w(m) z^{-m} = \sum_{m=-K}^{+K} w(m) z^{-m}. \qquad (A.1)$$

Evaluating $W(z)$ around the unit circle, i.e. substituting $z = e^{j\omega} = e^{j2\pi f}$ into Eq. (A.1), yields the frequency response

$$W(f) = \sum_{m=-K}^{+K} w(m) e^{-j2\pi m f}. \qquad (A.2)$$

The power density spectrum of this window is therefore

$$P(f) = |W(f)|^2 = W^*(f) W(f) \qquad (A.3)$$

where the asterisk superscript denotes complex conjugation. Substituting Eq. (A.2) into Eq. (A.3) yields

$$P(f) = \sum_{m_2=-K}^{+K} \sum_{m_1=-K}^{+K} w^*(m_2) e^{+j2\pi m_2 f} w(m_1) e^{-j2\pi m_1 f}$$

$$= \sum_{m_2=-K}^{+K} \sum_{m_1=-K}^{+K} w^*(m_2) w(m_1) e^{j2\pi(m_2-m_1)f}. \qquad (A.4)$$

The power over the interval $-f_\Delta \leq f \leq +f_\Delta$ is

$$P_{\Delta f} = \int_{-f_\Delta}^{+f_\Delta} P(f)\, df \qquad (A.5)$$

which may be represented using vector/matrix notation as

$$P_{\Delta f} = \mathbf{w}^\dagger \mathbf{Q}_{\Delta f} \mathbf{w} \qquad (A.6)$$

where the dagger superscript denotes a Hermitian transpose and $\mathbf{Q}_{\Delta f}$ is a $M \times M$ matrix, with the element in the $m_2$th row and $m_1$th column being the integral

$$Q_{\Delta f}^{m_2, m_1} = \int_{-f_\Delta}^{+f_\Delta} e^{j2\pi(m_2-m_1)f}\, df$$

$$= \frac{e^{j2\pi(m_2-m_1)f_\Delta} - e^{j2\pi(m_1-m_2)f_\Delta}}{j2\pi(m_2-m_1)} \qquad (A.7)$$

which are real numbers. The Slepian window maximizes the power concentration within the specified frequency interval. If the (band-limited) power over the interval $-f_\Delta \leq f \leq +f_\Delta$ is $P_{\Delta f}$ and the (total) power over the interval $-\frac{1}{2} \leq f \leq +\frac{1}{2}$ is $P_{1/2}$ then their ratio $\alpha$, with $0 \leq \alpha \leq 1$, is the Rayleigh quotient

$$\alpha = \frac{P_{\Delta f}}{P_{1/2}} = \frac{\mathbf{w}^\dagger \mathbf{Q}_{\Delta f} \mathbf{w}}{\mathbf{w}^\dagger \mathbf{Q}_{1/2} \mathbf{w}}. \qquad (A.8)$$

Due to the orthonormality of the complex sinusoids over the interval $-\frac{1}{2} \leq f \leq +\frac{1}{2}$, $\mathbf{Q}_{1/2} = \mathbf{I}$, where $\mathbf{I}$ is the identity matrix

$$\alpha = \frac{\mathbf{w}^\dagger \mathbf{Q}_{\Delta f} \mathbf{w}}{\mathbf{w}^\dagger \mathbf{w}}. \qquad (A.9)$$

The eigenvector corresponding to the greatest eigenvalue of $\mathbf{Q}_{\Delta f}$ maximizes this quotient and it is the Slepian window function for $m = -K \ldots +K$. Finally, a delay of $K$ samples is applied to yield a causal window for $m = 0 \ldots M-1$.

## Appendix B. Derivation of the Frequency-Domain Slepian Window

Like the time-domain window, the frequency-domain window maximizes the proportion of the power density response over the interval $-f_\Delta \leq f \leq +f_\Delta$ using a window that is finite and non-zero in the time domain for $m = -K \ldots +K$; however, in this case it is applied in the frequency domain using only $k = -B_{\text{win}} \ldots +B_{\text{win}}$ (see Section 2.2). This additional constraint reduces the achievable concentration but good responses are still possible, even for small $B_{\text{win}}$. The window is derived using the procedure described in Appendix A with a change of basis. The window is applied using

$$\hat{X}(n,k) = \sum_{\acute{k}=-B_{\text{win}}}^{+B_{\text{win}}} \tilde{w}(\acute{k}) \hat{X}_{\text{raw}}(n, k-\acute{k}). \tag{B.1}$$

The corresponding time-domain representation of this window is

$$w(m) = \sum_{k=-B_{\text{win}}}^{+B_{\text{win}}} \tilde{w}(k) e^{j2\pi mk/M} \tag{B.2}$$

which is represented more compactly in vector/matrix form using

$$\mathbf{w} = \mathbf{F}\tilde{\mathbf{w}} \tag{B.3}$$

where $\mathbf{F}$ is a $M \times (2B_{\text{win}} + 1)$ transformation matrix, with elements

$$F(m,k) = e^{j2\pi mk/M}. \tag{B.4}$$

Substituting Eq. (B.3) for $\mathbf{w}$ in Eq. (A.9) yields

$$\alpha = \frac{\tilde{\mathbf{w}}^\dagger \mathbf{F}^\dagger \mathbf{Q}_{\Delta f} \mathbf{F} \tilde{\mathbf{w}}}{\tilde{\mathbf{w}}^\dagger \mathbf{F}^\dagger \mathbf{F} \tilde{\mathbf{w}}} \tag{B.5}$$

which is solved as a generalized eigenvalue problem using $\mathbf{F}^\dagger \mathbf{F} = M\mathbf{I}$, due to the orthogonal columns of the time-to-frequency transformation. As in the time-domain case, the eigenvector corresponding to the greatest eigenvalue maximizes the concentration in the pass band and it is the Slepian window function for $m = -K \ldots +K$ and $k = -B_{\text{win}} \ldots +B_{\text{win}}$. The use of a non-causal window ensures that all elements of $\mathbf{F}^\dagger \mathbf{Q}_{\Delta f} \mathbf{F}$ are real, which simplifies its eigen-decomposition. The window must be delayed by $K$ samples in time domain to make it causal and so that its phase origin coincides with that of the analyzers (at $m = 0$). This is done via modulation with a complex sinusoid in the frequency domain using a multiplier of $e^{-j2\pi kK/M}$ for each bin of the window.

## Appendix C. Sum-of-Cosine Windows for Recursive DFTs

Sum-of-cosine windows are also applied using Eq. (B.1), where the elements $\tilde{w}(k)$ are simply arbitrary constants [28], with $\tilde{w}(0) = 1$ for proper normalization (0 dB gain at dc); for example $\tilde{w}(k) = [\frac{1}{2} \quad 1 \quad \frac{1}{2}]$, for $k = -1 \ldots +1$, for the Hann window. To shift the phase origin to $m = 0$ in the time domain, for the recursive filters considered in this paper, a factor of $e^{-j2\pi kK/M}$ is applied to the $k$th window coefficient. These windows are not optimal in any sense, although they are useful to lower side-lobes none-the-less. These windows benefit FIR and IIR DFT filters alike, provided the IIR poles are not too close to the unit circle. For IIR filters, this 'window function' applies a periodic modulation over an infinite extent in the time domain; however, the exponential decay of the associated filter's impulse response confines the window's influence to recent samples only.

**Appendix D. Derivation of the Stabilized IIR SDFT**

The 'mixing matrix' $H_{\text{mix}}$ is used in the frequency domain to orthonormalize a bank of recursive fading-memory filters. Orthonormalization of the impulse and frequency responses, ensures that

$$\sum_{m=0}^{\infty} h_{\text{ana},k_2}^*(m) h_{\text{ana},k_1}(m) = \delta_{k_1.k_2} \tag{D.1}$$

and

$$H_{\text{ana},k_1}(f) = H_{\text{ana},k_1}\left(\frac{k_2}{M}\right) = \delta_{k_1.k_2} \tag{D.2}$$

where $\delta$ is the Kronecker delta. The elements of $H_{\text{win}}$ are derived here using (D.1) which also ensures that Eq. (D.2) is satisfied.

Frequency analysis (and filtering) is interpreted here as a weighted least-squares fitting problem. It is assumed that

$$\tilde{x}(n-m) = \sum_{k=-B}^{+B} \beta_k(n) \psi_k^*(m) \tag{D.3a}$$

$$x(n) = \tilde{x}(n) + \varepsilon \tag{D.3b}$$

where $\psi_k(m) = e^{j\omega_k m}$ are the basis functions for the assumed sinusoidal signal model, $\beta_k(n)$ are the model coefficients at the time of the $n$th sample, $\tilde{x}(n)$ is the true signal value, $x(n)$ is the measured (noise-'corrupted') signal value, and $\varepsilon$ is additive zero-mean Gaussian noise $\varepsilon \sim \mathcal{N}(0, \sigma_x^2)$, with an unknown noise variance of $\sigma_x^2$. Maximum likelihood estimates $\hat{\beta}_k(n)$ of the parameters $\beta_k(n)$, derived using all past and present samples, are formed by minimizing the weighted sum-of-squared errors (SSE) where

$$\text{SSE}(n) = \sum_{m=0}^{\infty} \epsilon^*(n-m) w(m) \epsilon(n-m) \tag{D.4}$$

and where $\epsilon$ is the residual of the least-squares fit

$$\epsilon(n-m) = x(n-m) - \hat{x}(n-m) \tag{D.5}$$

and $w(m)$ is the error weighting function $w(m) = e^{\sigma m}$, with the forgetting-factor parameter $\sigma < 0$. In the context of our spectrum-estimation/signal-filtering problem, $\sigma$ determines the pole radius ($r = e^\sigma$) of the filter bank in the $z$ plane, $\psi_k(m)$ are the frequency components that determine the pole angle ($\omega_k = 2\pi k/M$) in the $z$ plane, and $\beta_k(n)$ is the frequency spectrum coefficient of the $k$th bin at the $n$th sample $\hat{X}(n,k)$. As the $\sigma, B$ and $K$ parameters (where $M = 2K+1$ and $B \leq K$) determine the pole positions of each first-order analyzer in the filter bank, this least-squares fitting process may be viewed as being an optimal zero-placement technique. If high-order analyzers (with non-unity pole multiplicity) are acceptable then $w(m) = m^\kappa e^{\sigma m}$ with $\kappa > 0$ may be used (see Section 2.5); however $\kappa = 0$ in what follows (to simplify the mathematical working and the computer coding).

If the least-squares fit is performed over a *finite* time interval $m = 0 \ldots M-1$, then the estimates of the model parameters $\beta_k$ (i.e. the spectrum coefficients) are determined in the usual way using

$$\widehat{\boldsymbol{\beta}} = \boldsymbol{\mathcal{B}}^{-1} \boldsymbol{\mathcal{A}} \boldsymbol{x}. \tag{D.6}$$

In Eq. (D.6): $\widehat{\boldsymbol{\beta}}$ is a column vector of length $2B+1$ containing the coefficient estimates $\widehat{\boldsymbol{\beta}} = [\hat{\beta}_{-B}, \ldots \hat{\beta}_{k_2}, \ldots \hat{\beta}_{+B}]^\dagger$; $\boldsymbol{x}$ is a column vector of length $M$ containing the recent input signal history $\mathbf{x} = [x(n), \ldots \ x(n-m), \ldots \ x(n-M+1)]^\dagger$; $\boldsymbol{\mathcal{A}} = \boldsymbol{\Psi}^\dagger \mathbf{W} \equiv H_{\text{ana}}$ and $\boldsymbol{\mathcal{B}} = \boldsymbol{\Psi}^\dagger \mathbf{W}\boldsymbol{\Psi} \equiv H_{\text{mix}}^{-1}$ where $\mathbf{W}$ is a square $M$ by $M$ matrix of zeros with the weighting vector $\mathbf{w} = [w(0), \ldots \ w(m), \ldots \ w(M-1)]$ along its diagonal, and $\boldsymbol{\Psi}$ is an $M$ by $2B+1$ matrix with the element in the $m$th row and $k_1$th column equal to $\psi_{k_1}^*(m)$.

If the least-squares fit is now performed over an *infinite* time interval $m = 0 \ldots \infty$, then the model parameters $\beta_{k_2}$ are also estimated using (D.6) however the finite summations in $\boldsymbol{\mathcal{B}}$ are now replaced by the infinite summations

$$\mathcal{B}_{k_2.k_1} = \sum_{m=0}^{\infty} \psi_{k_2}(m) w(m) \psi_{k_1}^*(m) \tag{D.7}$$

which may conveniently be evaluated in the $z$ domain using

$$\mathcal{B}_{k_2.k_1} = \mathcal{Z}\{\psi_{k_2}(m) w(m) \psi_{k_1}^*(m)\}\big|_{z=1}$$
$$= 1/\left[1 - e^{\sigma + j(\omega_{k_2} - \omega_{k_1})}\right]. \tag{D.8}$$

And taking the $\mathcal{Z}$ transform of $\boldsymbol{\mathcal{A}}$ yields the bank of analysis filters with the $k_1$th element being

$$\mathcal{A}_{k_1} = \mathcal{Z}\{w(m) \psi_{k_1}(m)\} = 1/[1 + az^{-1}] \tag{D.9}$$

where $a = -e^{\sigma + j\omega_{k_1}}$.

The mixing matrix and a Hann-like window may be applied in a single operation using $H_{\text{mix\&win}} = H_{\text{win}} H_{\text{mix}}$, where $H_{\text{win}}$ is a tri-diagonal matrix (for $B < K$) with the Hann window coefficients running along its diagonals (see Appendix C).